\documentclass[aps,prl,twocolumn,superscriptaddress,showpacs,showkeys,10pt]{revtex4-1}

\usepackage{amsmath}
\usepackage{graphicx}
\usepackage{color}

\DeclareMathOperator{\sech}{sech}

\newcommand{\ket}[1]{\left|#1\right>}

\begin{document}

\title{Qubit gates using hyperbolic secant pulses}

\author{H. S. Ku}\email{Hsiang-Sheng.Ku@nist.gov}
\affiliation{National Institute of Standards and Technology, Boulder, Colorado 80305, USA}
\author{J. L. Long}
\affiliation{National Institute of Standards and Technology, Boulder, Colorado 80305, USA}
\affiliation{Department of Physics, University of Colorado, Boulder, Colorado 80309, USA}
\author{X. Wu}
\affiliation{National Institute of Standards and Technology, Boulder, Colorado 80305, USA}
\author{M. Bal}
\affiliation{National Institute of Standards and Technology, Boulder, Colorado 80305, USA}
\author{R. E. Lake}
\affiliation{National Institute of Standards and Technology, Boulder, Colorado 80305, USA}
\author{Edwin Barnes}
\affiliation{Department of Physics, Virginia Tech, Blacksburg, Virginia 24061, USA}
\author{Sophia E. Economou}
\affiliation{Department of Physics, Virginia Tech, Blacksburg, Virginia 24061, USA}
\author{D. P. Pappas}
\affiliation{National Institute of Standards and Technology, Boulder, Colorado 80305, USA}

\date{\today}

\begin{abstract}
It has been known since the early days of quantum mechanics that hyperbolic secant pulses possess the unique property that they can perform cyclic evolution on two-level quantum systems independently of the pulse detuning. More recently, it was realized that they induce detuning-controlled phases without changing state populations. Here, we experimentally demonstrate the properties of hyperbolic secant pulses on superconducting transmon qubits and contrast them with the more commonly used Gaussian and square waves. We further show that these properties can be exploited to implement phase gates, nominally without exiting the computational subspace. This enables us to demonstrate the first microwave-driven $Z$-gates with a single control parameter, the detuning.
\end{abstract}

\pacs{03.67.Lx}

\maketitle

Controlled rotations of two-level systems were among the first examples of time-dependent quantum phenomena ever studied and continue to be a very active area of research owing to the central role they play in numerous quantum-based technologies currently being pursued \cite{Ladd_Nature10,Gisin_RMP02,Degen_arxiv16,Devoret_Science13}. Early investigations of two-level quantum dynamics were conducted using a double Stern-Gerlach experiment in which spins traverse a region of rotating magnetic field \cite{Phipps1932}. In the rotating frame, this problem can be mapped onto the familiar Rabi oscillations of atoms in a field \cite{Gerry}, where the drive strength, $\Omega(t)$, and detuning, $\Delta$, of the drive frequency from the two-level energy splitting correspond to the magnetic field strength and precession rate of the spins in the field. It was recognized early on by Rosen and Zener \cite{Rosen1932} that there is an exact solution to this problem, with $\Omega(t)= \Omega_0 \sech(\rho t)$. More importantly, for specific values of $\Omega_0$, a spin in an arbitrary superposition of $\ket{0}$ and $\ket{1}$ will always return back to that same initial state {\it independently} of the value of the detuning. This surprising result has been leveraged extensively in fields such as spatial solitons, quantum optics and self-induced transparency \cite{LambSolitons1971,SegevSpatialSolitons1992,McCall_PR69,LehtoSech2016}. The cyclic evolution is accompanied by the acquisition of opposite phases by states $\ket{0}$ and $\ket{1}$, which has been suggested \cite{Economou2006,Economou_PRL07,Economou2012,Economou2015,Barnes_arxiv16} as a means to implement phase gates or multi-qubit entangling gates in various qubit systems through the use of states outside the computational subspace.

Single-qubit gates are required for quantum computing and simulations. In the case of rotations about an axis in the $XY$~plane, the control design is rather straightforward: a resonant pulse of any shape will implement such a rotation, with the pulse area (the time integral of the pulse envelope) determining the angle of rotation. The theoretical simplicity of this concept has made qubit rotations about $XY$ axes routine in many labs \cite{Barends_Nature14,Johnson_NJP15,Blumoff_PRX16,Rol_arxiv16}. On the other hand, $Z$-rotations have been implemented to date via tuning \cite{SteffenTomographyScience2006}, combining rotations about $X$ and $Y$ using multiple pulses or just by keeping an accounting of all phases on the system \cite{McKayZ-GatesArxiv2017}. In general, these methods can result in increased decoherence in systems of multiple qubits, as it takes the parameters away from the high coherence regime. The other alternatives can result in increased overhead in pulse time or accounting. There is thus a need for a gate that achieves a rotation around the $Z$-axis using a single pulse to simplify and reduce overheads.

One of the difficulties in implementing $Z$-rotations (i.e. phase gates) is that, unlike in the case of rotations about axes in the $XY$ plane, there is no generic analytical solution for the evolution operator corresponding to a $Z$-rotation. One general requirement for phase gates is that the qubit undergoes a full Rabi flop, with all populations restored to their initial values. This can be achieved with resonant $2\pi$ pulses, however such pulses induce the same phase factor (equal to -1) to both the $|0\rangle$ and the $|1\rangle$ state, resulting in a trivial qubit operation that does not change the phase.

In the current work, we overcome this challenge and develop the first implementation of a microwave-mediated $Z$-gate, $Z_\mathrm{sech}(\Delta)$. By using a sech-function microwave pulse,  we are able to achieve a phase gate in qubits using only a single parameter, the detuning $\Delta$, and nominally driving the lower two levels of a transmon qubit. After fixing the peak strength of the pulse to satisfy the full Rabi flop condition, the detuning is used to tune the angle of rotation. Here we demonstrate the general result that this property is unique to the sech pulse, in contrast with other pulse shapes such as square and Gaussian, and demonstrate that we can use sech pulses to generate microwave-based phase gates that are intrinsically high-fidelity, $F>99~\%$.

To demonstrate a $Z$-gate with superconducting qubits, we use a microwave pulse with a $\sech$-envelope to rotate between the lowest two energy levels of a transmon. The drive pulse is defined as
\begin{equation}
\Omega(t)=\Omega_{0}\sech\left(\rho t\right)\cos{\left(\omega_\mathrm{D} t\right)},
\end{equation}
where $\Omega_{0}$ is the drive strength, $\rho$ is the pulse bandwidth, and $\omega_\mathrm{D}$ is the drive frequency. The full Rabi flop condition is satisfied by choosing
\begin{equation}
\Omega_0/\rho=n,
\end{equation}
where $n$ is an integer \cite{Rosen1932,McCall_PR69}. The salient feature is that this cyclic transition condition is independent of $\omega_\mathrm{D}$. This enables us to devise a single-control microwave $Z$-gate.

The $Z$-rotation is achieved as follows. Suppose the initial state of a qubit is in a superposition state \nolinebreak{$\Psi_0=a|0\rangle+b|1\rangle$}. After the qubit undergoes a cyclic evolution, the state $|0\rangle$ ($|1\rangle$) acquires a phase $\xi$ ($-\xi$), i.e., the state is driven to \nolinebreak{$\Psi=a|0\rangle+b\mathrm{e}^{i\phi}|1\rangle$} with $\phi=2\xi$. For a $2\pi$ pulse ($n=1$), the induced phase $\phi$ is given by
\begin{equation}
\phi=4\arctan\left(\rho/\Delta\right),
\label{Eq_phi}
\end{equation}
where the detuning is $\Delta=\omega_\mathrm{D}-\omega_{10}$ and $\omega_{10}$ is the transition frequency for the lowest two transmon levels. By fixing the drive strength $\Omega_0$ and the bandwidth $\rho$, we can construct a single-control $Z$-gate, $Z_\mathrm{sech}(\Delta)$, by adjusting only $\Delta$.

The specific device used for the experimental test of this gate is a concentric transmon \cite{Sandberg2013,Braumuller2016}. A transmon is essentially a nonlinear electrical LC oscillator that is read out using capacitive coupling, in the dispersive regime, to a linear resonator. The particular qubit used for this work had a transition frequency of $\omega_{10}=2\pi\times5.18$~GHz and an anharmonicity of $\omega_{10}-\omega_{21}=2\pi\times200$~MHz where $\omega_{21} $ is the transition frequency between the first and second excited states of the transmon. Further details of the qubit and the heterodyne readout method are given in the supplemental material.

\begin{figure}
\includegraphics[]{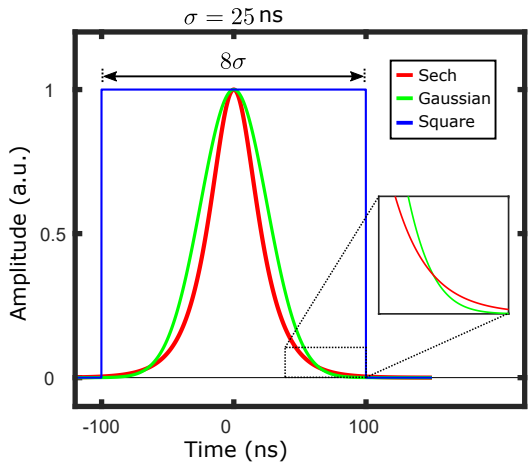}
\caption{(color online) Three different excitation profiles for sech-(red), Gaussian-(green), and square-(blue) pulse shapes.}
\label{fig_Pulse}
\end{figure}

In order to demonstrate the uniqueness of the sech-drive, Rabi oscillations were driven with sech-, Gaussian- and square-pulse envelopes as shown in Fig.~\ref{fig_Pulse}. The synthesized microwave drive is generated by using modulation signals from an arbitrary pulse sequencer to IQ modulate a local oscillator. The full length of the pulse extended over $\pm4\sigma$ in order to minimize sharp cut-offs of sech- and Gaussian-functions. The standard deviation $\sigma$ is related to the bandwidth $\rho$ by $\sigma=\pi/(2\rho)$. This range was chosen because it fully utilizes the full digitization range (8 bits) of the arbitrary pulse sequencer. As seen in the inset of Fig.~\ref{fig_Pulse}, the sech-pulse is slightly broader with a longer tail than the Gaussian. While improvements to the pulse shape could be made by either creating a hard on/off transition \cite{MartinisZgatePRA2014} or reducing leakage with DRAG \cite{Motzoi_PRL09,Kaufmann, McKayZ-GatesArxiv2017}, in this work we have chosen to use a simple sech-pulse shape for direct comparison to theory and general purpose applications.

\begin{figure}
 \includegraphics[]{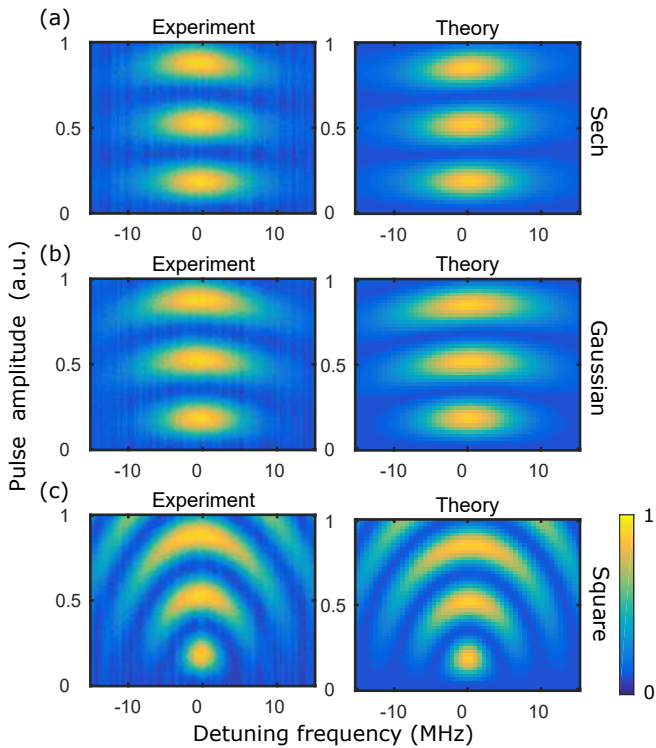}
 \caption{(color online) The excited state probability as a function of pulse amplitude (vertical axis) and detuning (horizontal) for (a) sech, (b) Gaussian, and (c) square pulses. The left and right panels compare experimental and theoretical simulations. The simulations impose the cutoff at $\pm4\sigma$ for the 8-bit digitization and include the four lowest energy levels of the transmon.}
\label{fig_RabiAmp2D}
\end{figure}

\begin{figure}
 \includegraphics{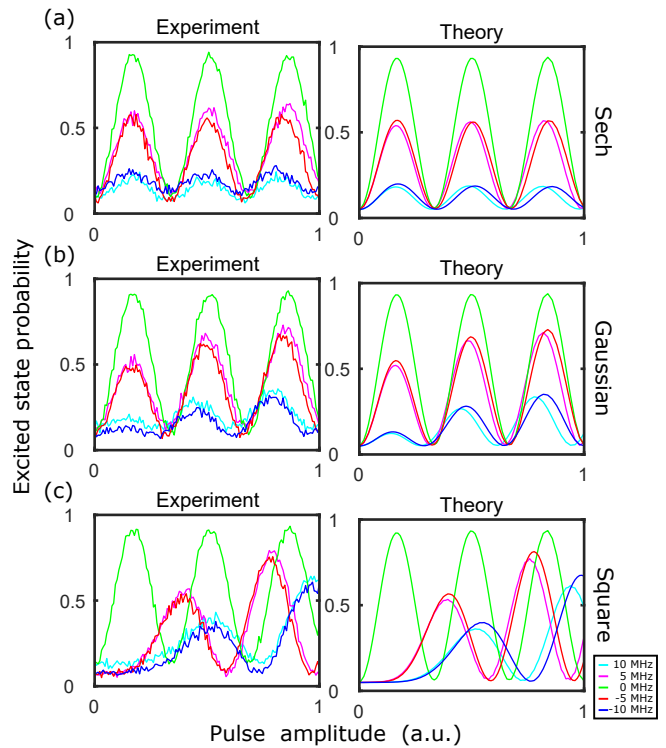}
 \caption{(color online) Line cuts at various detunings versus pulse amplitude from Fig.~\ref{fig_RabiAmp2D}.}
\label{fig_RabiAmp1D}
\end{figure}

\begin{figure}
 \includegraphics{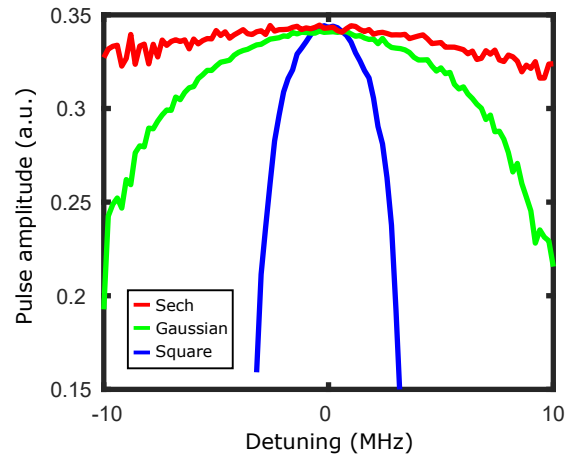}
 \caption{(color online) The pulse amplitude for the first full Rabi flop plotted versus the drive detuning for sech (red), Gaussian (green), and square (blue) pulse shapes.}
\label{fig_RabiCyclic}
\end{figure}

A comparison of experimental and theoretical Rabi oscillations versus the detuning $\Delta$ and the pulse amplitude are shown in Fig.~\ref{fig_RabiAmp2D} for $\sigma=25$~ns pulses. The theory is simulated using the empirical result that our qubit is typically initialized in an incoherently mixed state of 90~$\%$ $\left|0\right\rangle$ and 10~$\%$ $\left|1\right\rangle$ due to heating in the dilution refrigerator. The excited-state ellipses obtained from the sech pulses are qualitatively and quantitatively different compared to the chevron-shaped response exhibited by the Gaussian and square pulses. The first feature to note in the comparison is that the widths of the maxima, in the detuning axis, for the sech pulse are approximately the same for subsequent oscillation maxima [Fig.~\ref{fig_RabiAmp2D}(a)]. If we take cuts of the images along constant detunings, this leads to uniform periodic oscillations in the excited state probability as a function of pulse amplitude in the case of the sech, as shown in the 1-D plots of Fig.~\ref{fig_RabiAmp1D}(a). On the other hand, the Gaussian maxima in Fig.~\ref{fig_RabiAmp2D}(b) progressively widen and curve further downward with each oscillation period. The 1-D slices shown in Fig.~\ref{fig_RabiAmp1D}(b) for the Gaussian illustrate that the points where the population returns to the ground state shift toward lower pulse amplitude, while the Rabi oscillation contrast grows with increasing drive strength. This behavior is further exaggerated for the square pulses, as is evident in Fig.~\ref{fig_RabiAmp2D}(c) and Fig.~\ref{fig_RabiAmp1D}(c). The uniformity of the oscillations in the sech-pulse case is a direct reflection of the fact that, even if the drive amplitude is fixed, one can still achieve the cyclic condition. To see this effect quantitatively, the pulse amplitude for a full Rabi flop versus the drive detuning is plotted in Fig.~\ref{fig_RabiCyclic} for all three pulse shapes. The population return amplitudes are found by quadratic fits near the minimum region which corresponds to $2\pi$ pulses in Fig.~\ref{fig_RabiAmp2D}. The sech-pulse only has a $8.2~\%$ variation in pulse amplitude over the $\lvert\Delta\rvert\leq10$~MHz range, while the Gaussian-pulse has a $43.5~\%$ change. This low detuning-dependence behavior allows us to vary only one parameter, the detuning $\Delta$, to obtain an arbitrary $Z$-gate.

Although the experimental and theoretical results in Fig.~\ref{fig_RabiAmp2D}(a) are in very good agreement, we do see evidence of nonideality in the sech data from two sources. First, the finite bit resolution cutoff at $\pm{4}\sigma$ results in a flattening of the third oscillation. The second source is the existence of higher levels of the transmon system, resulting in a slight tilt of the Rabi ellipses. This behavior is illustrated in the theoretical simulation for multiple Rabi oscillations, and is observed to some extent in the experiment. However, note that only the first oscillation (about 1/3 of the $Y$-axis range in Fig.~\ref{fig_RabiAmp2D}(a)) is used to implement $Z$-gates, and for this region there is negligible discrepancy between theory and experiment. This high degree of frequency independence explains the very high fidelities we obtained, as we discuss below.

\begin{figure}
 \includegraphics[]{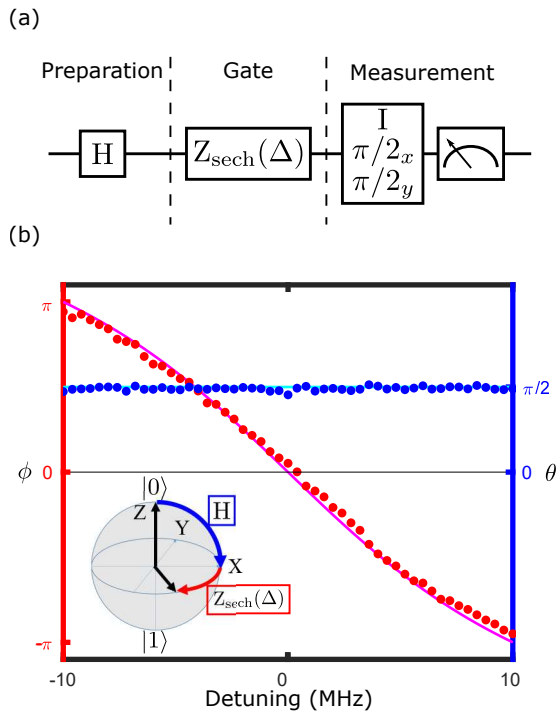}
 \caption{(color online) Tomography of a single-qubit phase gate, $Z_\mathrm{sech}(\Delta$), for a state prepared by a Hadamard gate. Panel (a) shows the pulse sequence and panel (b) shows experimental data (solid points) and simulations (solid curves) for the resulting angles $\theta$ and $\phi$.}
\label{fig_Zgate}
\end{figure}

The preparation and tomography pulse sequence for the $Z_\mathrm{sech}(\Delta$) phase gate is shown in Fig.~\ref{fig_Zgate}(a). In this example, the state is initially prepared by a $\pi/2$ rotation around the $Y$-axis, i.e., a Hadamard-like gate. A subsequent 2$\pi$-sech-pulse is then applied with the drive frequency given by $\omega_\mathrm{D}=\Delta+\omega_{10}$, and in each experimental run, either the $X$-, $Y$-, or $Z$-projection of the final state is measured to complete the single-qubit tomography. From this, the resulting $\phi(\Delta)$ and $\theta(\Delta)$ are determined, where $\theta$ is the angle of the Bloch vector from the $Z$-axis [solid points in Fig.~\ref{fig_Zgate}(b)]. For $-10~\mathrm{MHz}\leq\Delta\leq10~\mathrm{MHz}$, these data show that $\theta$ is constant at $\pi/2$, while $\phi$ is the angle of rotation of the Bloch vector around the $Z$-axis. Both are in excellent agreement with the prediction, Eq.~\eqref{Eq_phi} [solid lines in Fig.~\ref{fig_Zgate}(b)].

\begin{figure}
 \includegraphics[]{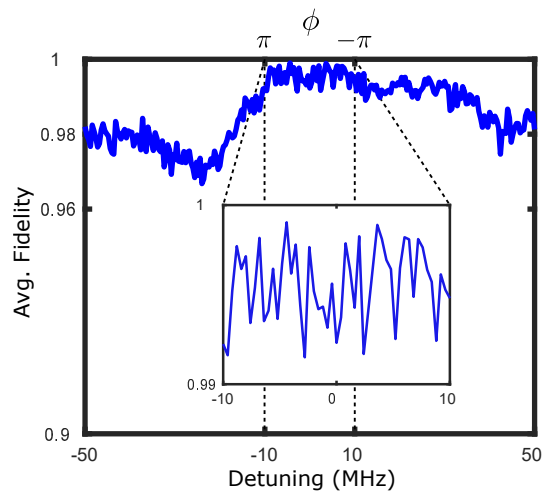}
 \caption{(color online) The average fidelity for the sequence in Fig.~\ref{fig_Zgate}(a) averaged over the six different initial states prepared as described in the text.}
\label{fig_AvgFid}
\end{figure}

To assess the performance of our $Z_\mathrm{sech}(\Delta)$ gate, we consider the fidelity of the rotations for the six input states obtained from $\pi/2$ rotations about the $\pm X$, $\pm Y$ directions, the identity operator, and a $\pi$ rotation. To account for the existence of mixed states in the initial state preparation, we calculate the fidelity of the gate operation, $Z_\mathrm{sech}(\Delta)$ as
\begin{equation}
F= \mathrm{Tr} \Bigg[ \sqrt{\sqrt{\rho_1}\rho_2\sqrt{\rho_1}} \Bigg],
\label{Fidelity}
\end{equation}
for each of the initial states. The density matrix $\rho_1$ is reconstructed from tomography measurements and $\rho_2$ is calculated from the theory, Eq.~\eqref{Eq_phi}, with a {9~\%} excited state before the state preparation pulse. As shown in Fig.~\ref{fig_AvgFid}, the fidelity averaged over the six different initial states is $F_\mathrm{avg}(\Delta)>99~\%$ for $\lvert\Delta\rvert\leq10$~MHz. This range corresponds to a $\mp\pi$ rotation around the Z-axis. For $\lvert\Delta\rvert\geq10$~MHz, there is a slight drop-off of $F_\mathrm{avg}$ for negative detuning, presumably due to the existence of  higher energy level transitions at lower frequencies due to the anharmonicity of the transmon.

In conclusion, we have shown that ideal sech-shaped pulses can be used to implement a fast, high-fidelity phase gate with a single control knob, the detuning. The unique properties of the sech allow us to achieve this gate while staying in the computational subspace throughout the duration of the gate. Our work paves the way toward high-fidelity two- and three-qubit entangling phase gates, which have been theoretically proposed based on the sech pulse \cite{Economou2015,Barnes_arxiv16} and can be advantageous when the lowest energy levels have some spread, as is intrinsic to the superconducting devices manufactured with lithographically defined and thermally oxidized components. Our results also lay the ground work for the superconducting circuit-based experimental demonstration of Self Induced Transparency \cite{McCall_PR69}, which occurs when, in addition to the temporal profile, the spatial distribution of the pulse is also a sech function. Such an experiment would be relevant for microwave-based logic with these circuits, slow-light demonstrations, and the development of larger scale circuits.

\begin{acknowledgments}
The authors acknowledge support of IARPA, the Army Research Office, and the NIST Quantum Initiative.
\end{acknowledgments}
\bibliography{sechbib}

\end{document}